\documentclass[10pt,letterpaper]{article} 

\usepackage[draft]{hyperref}
\usepackage{graphicx}
\usepackage{amsmath}
\usepackage{SIunits}
\usepackage{cite}
\begin{document}


\title{A Large Area Fiber Optic Gyroscope on multiplexed fiber network}
\author{C. Clivati$^{1,2}$, D. Calonico$^{1,*}$, G. A. Costanzo$^{2}$,\\ A. Mura$^1$, M. Pizzocaro$^{1,2}$ and F. Levi$^1$\\%
$^1$Istituto Nazionale di Ricerca Metrologica INRIM, \\strada delle Cacce 91, 10135, Torino, Italy\\
$^2$Politecnico di Torino,\\Corso Duca degli Abruzzi 24,10129, Torino, Italy \\
$^*$Corresponding author: d.calonico@inrim.it }
\maketitle

\begin{abstract}We describe a fiber optical gyroscope based on the Sagnac effect realized on a multiplexed telecom fiber network. Our loop encloses an area of \unit{20}{\kilo\metre\squared} and coexists with Internet data traffic. This Sagnac interferometer achieves a sensitivity of about (\unit{\power{10}{-8}}{\radian\per\second})$/\sqrt{\text{Hz}}$, thus approaching ring laser gyroscopes without using narrow-linewidth laser nor sophisticated optics. The proposed gyroscope is sensitive enough for seismic applications, opening new possibilities for this kind of optical fiber sensors.\end{abstract}



\noindent Optical sensing of ground rotations induced by earthquakes has been demonstrated with optical gyroscopes exploiting the Sagnac effect \cite{igel,belfi,pancha}. In particular, unlike other types of seismic sensors using an inertial mass as the reference, optical gyroscopes are not sensitive to translational motion. This makes these instruments very promising for understanding ground motion and field deployable rotation sensors based on the Sagnac effect have begun to be developed as a result \cite{velikoseltsev
}. Ring Laser Gyroscopes (RLGs) achieve a sensitivity well below \unit{\power{10}{-9}}{\radian\per\second}\cite{belfi, hurst, schreiber}; however these devices are not well suited for a massive or commercial implementation, as they require careful maintenance and sophisticated instrumentation such as narrow-linewidth lasers and complex optics. On the other hand, rotation sensors based on passive Fiber Optic Gyroscopes (FOGs) have a broader dynamic range, are transportable, and require only commercial components \cite{lefevre,culshaw}. However, ordinary FOGs are limited by shot noise, and not sensitive enough to measure rotational signals from distant earthquakes \cite{velikoseltsev}: their sensitivity typically ranges around \unit{\power{10}{-4}-\power{10}{-6}}{\radian\per\second}, with a single more relevant result in the \unit{\power{10}{-8}}{\radian\per\second} range \cite{jaroszewicz}. 

To exploit the advantages of FOGs while overcoming their sensitivity limitation, we realized a fiber gyroscope on a multiplexed telecom fiber enclosing a large urban area. The Sagnac interferometer proposed here achieves a sensitivity of about (\unit{\power{10}{-8}}{\radian\per\second})$/\sqrt{\text{Hz}}$, without using narrow-linewidth laser nor sophisticated optics; the setup is simple, uses off-the-shelf components and can be run on any optical telecom fiber loop enclosing an area. Therefore, devices such this one can be suitable for a distributed grid covering wide geographical regions. 

In this Letter we present the detailed setup of our FOG, the results in terms of rotational sensitivity, an analysis of the present limitations and a comparison with state-of-the-art RLGs. Then, we point out the possible use of this optical fiber sensor for seismic applications. 

\begin{figure}[htb]
\label{fig1}
\centerline{\includegraphics[width=7.5cm]{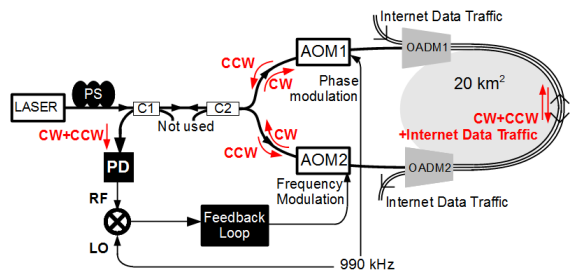}}
\caption{Setup of our FOG: CW (CCW) clockwise (counterclockwise) laser beam, PS polarization scrambler, C couplers, AOM acousto-optic modulators, PD photodiode, OADM Optical Add and Drop Multiplexers.\label{fig1}}
\end{figure}
In a Sagnac interferometer two laser beams counter propagate in an optical fiber loop enclosing an area, and accumulate a non-reciprocal phase shift \cite{lefevre}:
\begin{equation}
\label{eq1}
\varphi_\text{nr}=\frac{8\pi\nu}{c^2}\mathbf{A} \cdot \mathbf{\Omega},
\end{equation}
where $\mathbf{\nu}$ is the laser frequency, $c$ the 
speed of light in vacuum, $\mathbf{A}$ the area enclosed by the loop, and $\mathbf{\Omega}$ the rotation rate of the gyroscope reference frame. 

Our interferometer is composed by a \unit{47}{\kilo\metre} single mode commercial fiber located in the urban area around the city of 
Turin (Italy, colatitude \unit{45}{\degree}). Its shape is an elongated triangle with an enclosed area of \unit{20}{\kilo\metre\squared}, that is $\sim$1000 times larger than 
the uppermost values achieved in state-of-the-art FOGs \cite{jaroszewicz}. The resulting phase due to the Earth rotation is thus \unit{55}{\radian}.\\ 
The optical fiber is used for 
the Internet data traffic and is implemented on a Dense Wavelength Division Multiplexing (DWDM) architecture, with $\sim$\unit{23}{\deci\bel} of optical losses. The optical signal is a laser radiation at \unit{1542}{\nano\metre}, corresponding to the 44th channel of the International Telecommunication Union (ITU) grid, while Internet data are transmitted on the 21st and 22nd channel, \unit{2}{\tera\hertz} away. There is no evidence of any crosstalk between the channels. In Figure \ref{fig1} the experimental setup is shown, in a minimum configuration scheme: laser radiation at \unit{1542}{\nano\metre} provided by a fiber laser is injected in the interferometer and split into two beams travelling over the loop in opposite directions, the first clockwise (CW), and the second counterclockwise (CCW). The setup design assures that the two beams travel exactly the same fiber. The polarization of the injected light is randomized through a polarization scrambler PS; this reduces the effect of polarization mode dispersion along the fiber, i. e. a source of phase shifts between CW and CCW beams. Scrambling the polarization also reduces the influence of backscattering: in fact, it adds some phase noise on the optical carrier, shortening the coherence time of the radiation. Before being coupled into the urban fiber loop, the two beams are frequency shifted by \unit{40}{\mega\hertz} with two acousto-optic modulators (AOMs). AOM1 also modulates the phase of the optical carrier at a rate of $f_\text{m}=\unit{990}{\kilo\hertz}$: the CW beam is thus phase modulated immediately at the beginning of the loop, whereas the CCW beam is modulated after a round trip. The two beams, with optical power of \unit{3}{\milli\watt} each, are injected in the telecom fiber loop using the Optical Add and Drop Multiplexers OADM1 and OADM2. While travelling over the fiber in opposite directions, CW and CCW beams accumulate a phase difference $\varphi_\text{nr}$ due to the Sagnac effect. After a round trip, they are extracted from the telecom loop and recombined on the photodiode PD. Since there is a time delay $\tau$=\unit{235{\micro}\second} between the phase modulation of the two beams, the current from the photodiode also varies at the phase modulation rate $f_\text{m}$. It can be expressed in the form \cite{ezekiel}
\begin{align}
\label{eq2}
I =I_0&+I_1J_1(x)\sin\varphi_\text{nr} \cos\Big[2\pi f_\text{m}\big(t-\frac{\tau}{2}\big)\Big]+\nonumber \\
&+ f_\text{m} \, \text{harmonics}
\end{align}
$I_0$ and $I_1$ are respectively the amplitudes of the dc and the first harmonic signal and depend on the optical power of the beams, $J_1$ is the first Bessel function of the first kind, $x=2\phi_0 \sin (2\pi f_\text{m} \tau)$, and $\phi_0$ is the phase deviation depth. $f_\text{m}$ and $\phi_0$ are set to maximize the first harmonic term. This signal is processed to extract $\varphi_\text{nr}$, i. e. the phase due to the Sagnac effect. This requires a closed-loop system, in which $\varphi_\text{nr}$ is compensated by a frequency offset $\Delta_f$  between the two beams \cite{ezekiel}. When the loop is closed, $\Delta_f$ satisfies the relation
\begin{equation}
\label{eq3}
\varphi_\text{nr}\pm 2\pi k=\frac{2 \pi n L}{c} \Delta_f
\end{equation}
where $n$ is the refractive index of the optical fiber, $L$ the loop length, and $k$ an integer. AOM2 is used as the actuator of the feedback loop. $\Delta_f$ is set below \unit{100}{\hertz} to reduce the sensitivity on optical path length variations $\delta(nL)$, that can induce a non-reciprocal phase $\varphi=(2 \pi /c ) \Delta_f \delta(nL)$.
Assuming that $\varphi_\text{nr}$ is only due to the Sagnac effect, from eq. (\ref{eq1}) and eq. (\ref{eq3}):
\begin{equation}
\label{eq4} 
\Delta_f=\frac{4\nu}{nLc}\mathbf{A} \cdot \mathbf{\Omega}
\end{equation}
Thus, the correction frequency can be recorded to extract information about $\Omega$. This formula is the same as for RLGs, but the derivation is different.  

The noise spectral density of the frequency correction signal was acquired with a Fast Fourier Transform Spectrum Analyzer and is shown in Figure \ref{fig2} (left-hand axis). The frequency correction signal was converted into the rotation rate using eq. (\ref{eq3}); the scaled result is shown on the right-hand axis. The Sagnac phase shift depends in principle on tilt variations (change in $\mathbf{\Omega}/\Omega$) and spin variations (change in $\Omega$); however, the sensitivity to variations of tilt is about $\sim 10^4$ times lower than the sensitivity to variations of spin.
Figure \ref{fig2} also shows the present sensitivity limit of this FOG, and, for comparison, the sensitivity of three RLGs: G-Pisa, in Italy \cite{belfi}, UG-II, in New Zealand \cite{hurst}, and G, in Germany \cite{schreiber}. 
\begin{figure}[htb]
\label{fig2}
\centerline{\includegraphics[width=7.5cm]{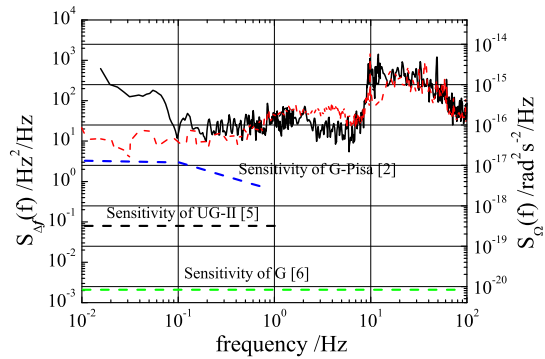}}
\caption{Power spectral density of the measured frequency (black solid line, left y axis) and equivalent spin variations (right y axis). Red line: contribution from fiber's mechanical noise; dashed lines: sensitivity of G-Pisa (blue)  \cite{belfi}, UG-II (black)  \cite{hurst}, G (green)  \cite{schreiber}.\label{fig2} 
}
\end{figure}

The sensitivity of our setup is mainly limited by nonreciprocal optical length variations. When travelling along the loop, the optical carrier acquires a phase noise $S_{\varphi_\text{F}}(f)$ due to the mechanical noise of the fiber \cite{williams}. Our interferometer is at first order insensitive to this noise, since CW and CCW beams travel in the same fiber; however, since they travel in opposite directions, the uncorrelated residual optical length variation leads to a nonreciprocal phase noise. Following an approach similar to \cite{williams}, we can estimate this contribution to be 
\begin{equation}
\label{eq5}
S_{\varphi_\text{F,NR}}(f)=\frac{1}{3}(2\pi f \tau)^2 S_{\varphi_\text{F}}(f)
\end{equation}
$S_{\varphi_\text{F}}(f)$ was estimated from previous measurements on the fiber loop used \cite{calonico} and it is the ultimate limitation to the sensitivity of our setup. Figure \ref{fig2} shows the equivalent frequency correction due to this noise source. This contribution merely depends on the loop length: thus for a given fiber length, it is beneficial to maximize the enclosed area. For instance, in our configuration, keeping the same \unit{20}{\kilo\metre\squared} area, the sensitivity could have been improved of a factor $\sim26$ if the loop had not been an elongated triangle but a circle (i.e. \unit{15.8}{\kilo\metre} length). In addition, $S_{\varphi_\text{F}}(f)$ scales as $L$ and $S_{\varphi_\text{F,NR}}(f)$ scales as $L^3$ \cite{williams}, whereas the Sagnac phase spectral density scales as $A^2 \sim L^4$, thus the sensitivity could be further improved for fiber loops enclosing a wider area.  

Minor sensitivity contributions could come from the Kerr effect and scale factor instability, i.e. variations in the area $A$ or length $L$. The phase dependence on the optical power as predicted by the Kerr effect \cite{agrawal} was assessed changing the optical power of the CW beam with respect to the CCW. Even with a power unbalance of 20\%, Kerr related effects were below the sensitivity, so this contribution is negligible for optical power fluctuation in normal operation, i. e. less than 5\%. Concerning the scale factor instability, the sensitivity of our setup to length variation is \unit{0.8}{\hertz\per\metre}, whereas the sensitivity to area variations is 2$\times$\unit{\power{ 10}{-3}}{\hertz\per\metre\squared}. Therefore, for any reasonable variation of $L$ or $A$, the phase contribution from these noise sources is well below our reported sensitivity.  

The instability over long measuring times has been evaluated through the Allan deviation $\sigma_{\Delta_f}(\tau)$ of the correction frequency $\Delta_f$ as a function of the averaging time $\tau$, for a dataset of about 3 days of uninterrupted measurement. Figure \ref{fig3} shows both the Allan deviation of the frequency and of the corresponding rotation signal $\sigma_{\Omega}(\tau)$.
\begin{figure}[htb]
\label{fig3}
\centerline{\includegraphics[width=7.5cm]{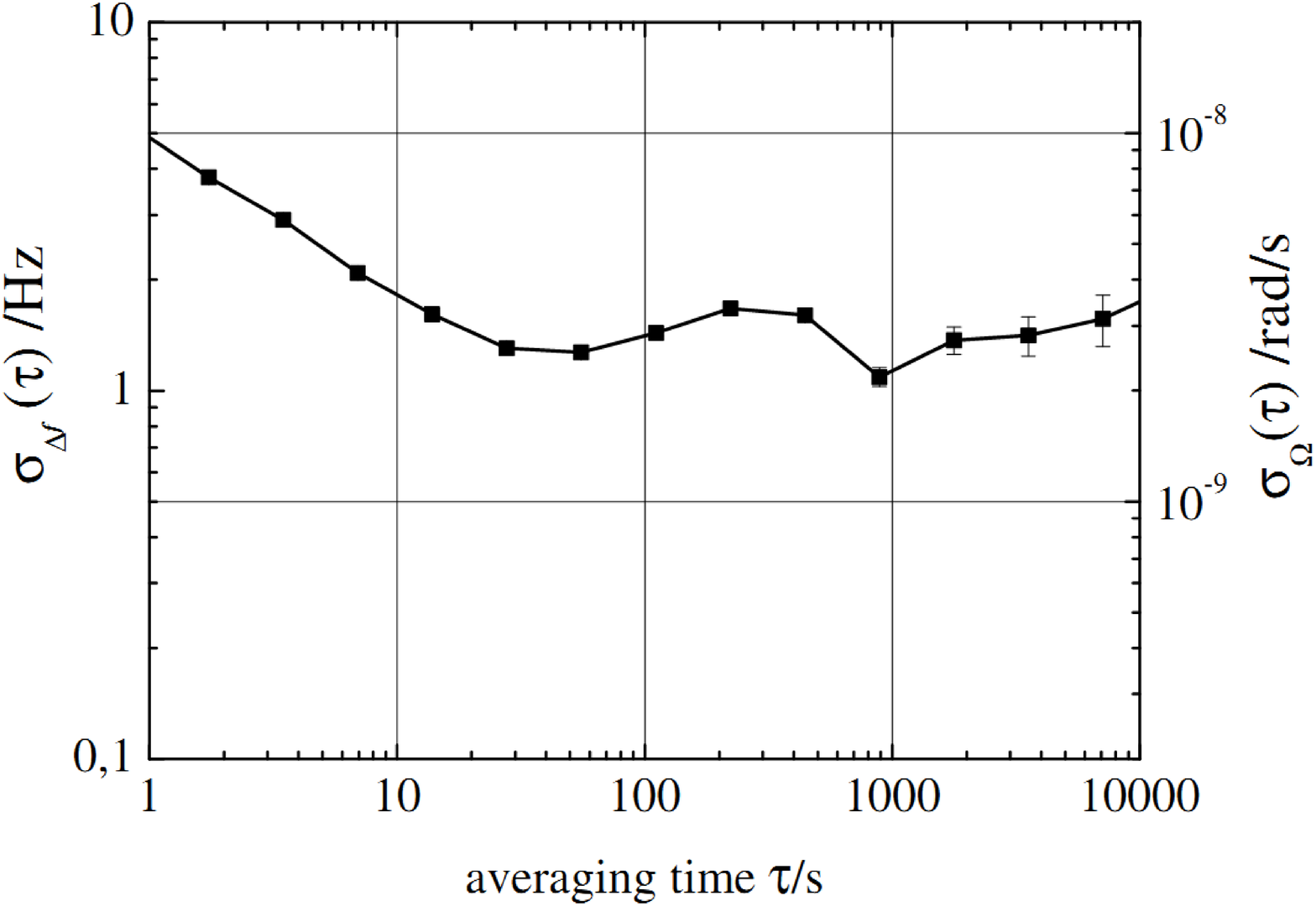}}
\caption{Allan deviation of the correction frequency $\sigma_{\Delta_f}(\tau)$ (left-hand y axis) and the equivalent instability $\sigma_{\Omega}(\tau)$ of the spin variations (right-hand y axis).\label{fig3} }
\end{figure}
The excess of instability on timescales longer than \unit{100}{\second} is due to incomplete polarization scrambling, that makes our system still sensitive to long term polarization drifts. This effect will be reduced by improving the depolarization stage.  

In conclusion, we have demonstrated that ordinary optical fiber networks, used for Internet data traffic, could be exploited to implement optical rotation sensors based on the Sagnac effect. This experiment could pave the way for a possible network of these sensors for seismic detection, placed side by side with traditional instrumentation, supported by the increasing activity on fiber networks for frequency metrology \cite{schnatz,lopez} and references therein. The advantage of the proposed setup is a good sensitivity, obtained using wide infrastructures already commercially available, such as the optical fiber networks, without the development of sophisticated experiments such a RLG. The sensitivity is not yet fully exploited, and improvements beyond the demonstrated performances are feasible. This type of rotation sensor could offer new opportunities for detection of seismic events, provided a deeper investigation on some issues. These include: a better understanding on how the groung motion is detected by a large-area sensor, and the feasibility of a real distributed grid of gyroscopes. These issues are currently under study. \\

The authors acknowledge  N. Beverini and J. Belfi for careful reading of the manuscript; G. Carelli and L. Sambuelli for useful discussions; the GARR Consortium for technical help with the fiber infrastructure  and Compagnia di San Paolo for funding.

\bigskip

\pagebreak
\begin{enumerate}
\item{H. Igel, K. U. Schreiber, A. Flaws, B. Schuberth, A. Velikoseltsev, A. Cochard, ``Rotational motions induced by the M 8.1 Tokachi-oki earthquake, September 25, 2003'', Geophis. Res. Lett. \textbf{32}, L08309 (2005).}
\item{J. Belfi, N. Beverini, G. Carelli, A. Di Virgilio, E. Maccioni, G. Saccorotti, F. Stefani, A. Velikoseltsev, ``Horizontal rotation signals detected by ``G-Pisa'' ring laser for the M=9.0, March 2011, Japan earthquake'', J Seismol \textbf{16}, 767 (2012).}
\item{A. Pancha, T. H . Webb, G . E. Stedman, D . P. McLeod, and K. U. Schreiber, ``Ring laser detection of rotations from teleseismic waves'', Geophys. Res. Lett. \textbf{27}, 3553 (2000).}
\item{A. Velikoseltsev, K. U. Schreiber, A. Yankovsky, J.-P. R. Wells, A. Boronachin, A. Tkachenko, ``On the application of fiber optic gyroscopes for detectionof seismic rotations`` J. Seismol. \textbf{16}, 623 (2012).}
\item{R. B. Hurst, G. E. Stedman, K. U. Schreiber, R. J. Thirkettle, R. D. Graham, N. Rabeendran, and J.-P. R. Wells, ``Experiments with an \unit{834}{\metre\squared} ring laser interferometer'', J. Appl. Phys \textbf{105}, 113115 (2009).} 
\item{K.U. Schreiber, A. Velikoseltsev, G. E. Stedman, R. B. Hurst, T. Klugel}{New applications of very large ring lasers, in \textit{Symposium Gyro Technology}, p. 8.0 (Sorg H, ed. 2003).}
\item{H. Lefevre ``\textit{The fiber-optic gyroscope}'' (Artec House, 2003).}
\item{B. Culshaw, ``The optical fibre Sagnac interferometer: an overview of its principles and applications'', Meas. Sci. Technol. \textbf{17}, pp. R1-R16, (2006).}
\item{L. R. Jaroszewicz, Z. Krajewski, H. Kowalski,G. Mazur, P. Zinowko, J. Kowalski, ``AFORS: Autonomous Fibre-Optic Rotational Seismog raph: Design and Application'', Acta Geophysica \textbf{59}, 578 (2011).} 
\item{J. L. Davis, S. Ezekiel, ``Closed-loop, low-noise  fiber-optic rotation sensor'', Opt. Lett. \textbf{6}, 505 (1981).}
\item{W. Williams, W. C. Swann, and N. R. Newbury, ``High-stability transfer of an optical frequency over long fiber optic links,'' J. Opt. Soc. Am. B \textbf{25}, 1284 (2008).}
\item{D. Calonico, C. Clivati, G. A. Costanzo, A. Godone, F. Levi, M. Marchetti,  A. Mura, M. Prevedelli, M. Schioppo, G. M. Tino,  M. E. Zucco, N. Poli, ``Optical Frequency Link between Torino and Firenze for remote comparison between Yb and Sr optical clocks'' in \textit{proceedings of the EFTF} (2012), p. 396.}
\item{G. Agrawal, ``\textit{Nonlinear Fiber Optics}'' 3rd ed. (Academic, 2001).}
\bibitem{schnatz}{K. Predehl, G. Grosche, S. M. F. Raupach, S. Droste, O. Terra, J. Alnis, Th. Legero, T. W. Hansch, Th. Udem, R. Holzwarth, H. Schnatz, ``A 920-Kilometer Optical Fiber Link for Frequency Metrology at the 19th decimal place'', Science \textbf{336}, 441, (2012).}
\bibitem{lopez}{O. Lopez, A. Haboucha, F. Kefelian, H. Jiang, B. Chanteau, V. Roncin, C. Chardonnet, A. Amy-Klein, G. Santarelli, ``Cascaded multiplexed optical link on a telecommunication network for frequency dissemination'', Opt. Expr. \textbf{18}, 16849 (2010).}


\end{enumerate}

\end{document}